\documentclass[prd,twocolumn,showpacs,nofootinbib]{revtex4}
\usepackage{times}
\usepackage{natbib}
\usepackage{epsfig}

\newcommand{\apjl}{Astrophys. J. Lett.}
\newcommand{\apjs}{Astrophys. J. Suppl. Ser.}
\newcommand{\aap}{Astron. Astrophys.}

\newcommand{\araa}{Annu. Rev. Astron. Astrophys.}
\newcommand{\physrep}{Phys. Rep.}

\newcommand{\ang}{\hat\nabla}

\newcommand{\bdv}[1]{{\bf #1}}

\newcommand{\up}[1]{{\rm #1}}
\newcommand{\kms}{{\rm km\, s}^{-1}}
\newcommand{\mpc}{{\rm Mpc}}
\newcommand{\hmpc}{{h^{-1}\mpc}}
\newcommand{\msun}{M_{\odot}}
\newcommand{\hmsun}{{h^{-1}\msun}}
\newcommand{\Rvir}{R_\up{vir}}
\newcommand{\OM}{\Omega_m}
\newcommand{\OB}{\Omega_b}
\newcommand{\rms}{\sigma_8}

\newcommand{\vL}{\bdv{l}}
\newcommand{\Vang}{\bdv{\hat n}}
\newcommand{\Vangm}{\bdv{\hat m}}
\newcommand{\lenT}{\tilde T}
\newcommand{\spix}{\sigma_\up{pix}}
\newcommand{\Npix}{N_\up{pix}}
\newcommand{\fsky}{f_\up{sky}}
\newcommand{\Tobs}{\tilde T^\up{obs}}

\newcommand{\That}{\hat T}
\newcommand{\lenC}{\tilde C}
\newcommand{\Cobs}{\tilde C^\up{obs}}
\newcommand{\sbeam}{\sigma_b}
\newcommand{\tfwhm}{\theta_\up{FWHM}}
\newcommand{\khat}{\hat\kappa}
\newcommand{\Cphi}{C^{\phi\phi}}
\newcommand{\Ckap}{C^{\kappa\kappa}}
\newcommand{\Cdd}{C^{dd}}
\newcommand{\Nkap}{N^{\kappa\kappa}}
\newcommand{\tE}{\theta_\up{E}}
\newcommand{\tEm}{\theta_\up{E}^m}
\newcommand{\tEms}{\theta_\up{E}^{m2}}
\newcommand{\tEst}{\theta_\up{E}^\star}

\newcommand{\tEQE}{\hat\tE^\up{QE}}
\newcommand{\tEML}{\hat\tE^\up{ML}}
\newcommand{\tEopt}{\hat\tE^\up{opt}}
\newcommand{\beeq}{\vspace{10pt}\begin{equation}} 
\newcommand{\eneq}{\vspace{10pt}\end{equation}}
\newcommand{\bear}{\vspace{10pt}\begin{eqnarray}}
\newcommand{\enar}{\end{eqnarray}{\\\\ \noindent}}
\newcommand{\Fisher}{\mathcal{F}}
\newcommand{\Lik}{\mathcal{L}}
\newcommand{\lcut}{l_\up{cut}}
\newcommand{\muK}{\mu\up{K}}
\newcommand{\Minit}{M_\up{init}}
\newcommand{\bhat}{\bdv{\hat s}}
\newcommand{\dsmall}{\bdv{\hat\Delta}}

\begin{document}

\title{Improved estimation of cluster mass profiles 
from the cosmic microwave background}

\author{Jaiyul Yoo$^1$}
\altaffiliation{Electronic address: jyoo@cfa.harvard.edu} 
\author{Matias Zaldarriaga$^{1,2}$}
\affiliation{$^1$Harvard-Smithsonian Center for Astrophysics, Harvard 
University, 60 Garden Street, Cambridge, MA 02138}
\affiliation{$^2$Jefferson Physical Laboratory, Harvard University, 
17 Oxford Street, Cambridge, MA 02138}

\begin{abstract}
We develop a new method for reconstructing cluster mass profiles and 
large-scale structure from the cosmic microwave background (CMB). By analyzing
the likelihood of CMB lensing, we analytically prove that standard quadratic
estimators for CMB lensing are unbiased and achieve the optimal condition
only in the limit of no lensing; they become progressively biased and
sub-optimal, when the lensing effect is large, especially for clusters that
can be found by ongoing Sunyaev-Zel'dovich surveys. Adopting an alternative
approach to the CMB likelihood, we construct a new maximum likelihood 
estimator that utilizes delensed CMB temperature fields based on an assumed
model. We analytically show that this estimator asymptotically approaches
the optimal condition as our assumed model is refined, and we numerically
show that as we iteratively apply it to CMB maps
our estimator quickly converges to the true model with a factor of ten
less number of clusters than standard quadratic estimators need.
For realistic CMB experiments, we demonstrate the 
applicability of the maximum likelihood estimator with tests against
numerical simulations in the presence of CMB secondary contaminants. With 
significant improvement on the signal-to-noise ratio, our new maximum 
likelihood estimator can be used to measure the cluster-mass 
cross-correlation functions at different redshifts, probing the evolution
of dark energy. 
\end{abstract}


\maketitle

\section{Introduction}
\label{sec:intro}
As the most distant observable sources, the cosmic microwave background (CMB)
anisotropies provide a unique channel to probe the universe after the 
cosmological recombination epoch. In particular, weak gravitational lensing of 
the CMB can be used to map the matter distribution in the universe at higher 
redshift than weak lensing of faint background galaxies can ever achieve. 
Recent work \citep{HIPAET04,SMZADO07,HIHOET08,CASLET08}
has focused on measuring the lensing signature in the CMB
by large-scale structure between the last scattering surface and the present
universe, but relatively little attention has been paid to weak lensing 
of the CMB by clusters of galaxies.

The abundance of massive clusters is exponentially sensitive to the growth of 
the underlying matter distribution, and hence it has been recognized as a
powerful probe of the evolution of dark energy (e.g., \citep{DETF06}).
However, the constraining power as a cosmological probe
can be only realized, if the cluster masses are accurately measured.
To achieve this goal, many cluster surveys are designed to detect massive
clusters and measure their mass using the Sunyaev-Zel'dovich (SZ)
effect, and some of the planned
surveys are already operational using the South Pole Telescope (SPT),
and the Atacama Cosmology Telescope (ACT).
Weak lensing of the CMB can be applied to the same clusters found in the
SZ surveys without additional observations,
providing independent measurements of their mass. 
Furthermore, the CMB provides the highest redshift source plane with precision 
measurements of its distance, which can be combined with galaxy weak lensing 
measurements of the same lensing clusters to obtain angular diameter 
distance ratio estimates that are independent of the mass distribution,
substantially increasing the leverage to constrain cosmological
parameters \citep{HUHOVA07}.

Gravitational lensing by clusters imprints a unique signature in the CMB 
anisotropies. On arcminute scales, the primordial CMB anisotropies decay
exponentially due to the photon diffusion from the baryon-photon fluid around
the recombination epoch \citep{SILK68},
and to a good approximation the CMB can be considered as a pure temperature
gradient 
on small scales. Based on this approximation, \citet{SEZA95} showed that 
clusters create dipole-like wiggles in the CMB temperature by remapping the
otherwise smooth gradient field,
and this unique feature can be used to isolate the lensing effect
by clusters and to reconstruct the deflection angle, once the temperature
gradient is separately measured on large scales.
\citet*{VAAMWH04} and \citet{HOKO04} used $N$-body simulations 
to model realistic lensing clusters, and they found that 
the mass reconstruction for individual clusters is compromised,
since it is hard to measure the large-scale temperature gradient accurately
and secondary anisotropies in the CMB can partially 
mimic the lensing signature. 

However, it has been realized that one can apply the same technique developed 
for reconstructing large-scale structure to clusters of galaxies, measuring
the statistical properties of a sample of clusters.
Unlike galaxy weak lensing, CMB anisotropies have no characteristic shape, 
even statistically, from which the deviation is a measure of the lensing 
effect. Gravitational lensing, however, gives rise to a deviation 
of the two-point correlation function of the CMB temperature anisotropies
from statistical isotropy. 
The standard technique is to construct a lensing estimator that is quadratic in 
observed temperature anisotropies, measuring the  
correlation between different Fourier modes, which is directly proportional
to the lensing effect \citep{HU01b}.

This method is easy to implement in analyzing real data compared to the full
likelihood analysis \citep{HISE03a} and no separate measurement is required
to obtain the large-scale temperature gradient. However,
\citet{MABAMEET05} showed that standard quadratic estimators need a 
modification to be an unbiased estimator in a region around massive clusters.
\citet*{HUDEVA07} quantitatively demonstrated that
standard quadratic estimators based on the linear approximation ignore
 higher-order terms in the lensing effect that coherently contribute to
the lensing reconstruction, and hence the reconstruction is biased low when 
the lensing effect is large. Furthermore, they proposed modified quadratic 
estimators that remove the higher-order terms in violation of the linear
approximation by low-pass filtering observed temperature fields, and they
showed that the modified quadratic estimators recover cluster mass profiles
with no significant bias.
However, the cutoff scale of the low-pass filter is somewhat arbitrary and
it depends on the lensing effect, which we want to measure with the estimators.

Here we develop a new maximum likelihood estimator for reconstructing cluster
mass profiles and large-scale structure by analyzing the likelihood of CMB
lensing. Our approach is similar in making full use of the likelihood 
information to one advocated by \citet{HISE03a}. While they derive an 
analytic expression for a maximum likelihood estimator, it is impractical to
apply to a realistic problem, because the solution is too general and 
computationally expensive. However, our maximum likelihood estimator is 
different from theirs and it is easy
to use in practice, because we adopt an alternative approach to setting up
the likelihood: it takes a similar form of standard quadratic estimators and it 
approaches the optimal condition as it is iteratively applied to CMB maps.
Furthermore, we show that our maximum likelihood estimator can reconstruct
cluster mass profiles with a factor of ten less number of clusters than 
standard or modified quadratic estimators need.

The rest of the paper is organized as follows.
We first derive a quadratic
estimator, accounting for the telescope beam effect
in Sec.~\ref{sec:formalism}. This consideration
makes a difference compared to the usual practice in the literature,
where quadratic estimators are 
often applied to beam deconvolved CMB maps.
In Sec.~\ref{sec:mle} we analytically show that the
quadratic estimators are unbiased and optimal only when the lensing effect 
vanishes, and why the modified quadratic estimators outperform the 
standard quadratic 
estimators when the lensing effect is large. Based on this observation, we 
construct a delensed temperature field and derive a maximum likelihood 
estimator using the delensed temperature field. 
We demonstrate its applicability to realistic CMB experiments using numerical
simulations in Sec.~\ref{sec:num}. 
We discuss the impact
of the telescope beam and instrumental noise in the delensing process
and we conclude in Sec.~\ref{sec:con}.

In this paper we will only
consider lensing estimators based on CMB temperature
anisotropies, since the planned surveys are not yet sensitive to CMB
polarization anisotropies on arcminute scales.
However, it is straightforward to extend our formalism
to lensing estimators based on CMB polarization anisotropies. 
Throughout the paper
we assume a flat $\Lambda$CDM universe with the matter density 
parameter $\OM h^2=0.127$, the baryon density parameter $\OB h^2=0.0222$, the 
Hubble constant $h=0.73$, the spectral index $n_s=0.95$, the optical depth to
the last scattering surface $\tau=0.09$, and the primordial
curvature perturbation amplitude $A_s=2.5\times10^{-9}$ (corresponding to the
matter power spectrum normalization $\rms=0.75$), consistent with the recent
cosmological parameter estimation (e.g., \citep{TESTET06,SPBEET07,KODUET08})

\section{Formalism}
\label{sec:formalism}
Here we describe our notations for weak lensing of the
CMB and derive a quadratic estimator for CMB lensing reconstruction.

\subsection{Weak Lensing of the CMB}
Gravitational lensing deflects light rays as they propagate through
fluctuating gravitational fields, and the deflection vector 
$\bdv{d}(\Vang)$ at the angular position $\Vang$ on the sky is related to
the line-of-sight projection of the gravitational potential $\psi$ as
$\bdv{d}(\Vang)=\bdv{\ang}\phi(\Vang)$, where the projected potential is
\beeq
\phi(\Vang)=-2\int_0^{D_\star}\!\! 
dD~{D_\star-D\over D D_\star}~\psi(D\Vang,D),
\eneq
$\ang$ is the derivative with respect to $\Vang$, 
and $D_\star$ is the comoving angular diameter distance to the last scattering
surface. Here we have assumed a flat universe and $c\equiv1$. 
The projected potential is further
related to the convergence $\kappa$ as $\ang^2\phi(\Vang)=-2\kappa(\Vang)$.

Since gravitational lensing conserves the surface brightness of diffuse 
backgrounds, the lensed temperature field $\lenT(\Vang)$ 
of the CMB is simply the intrinsic (unlensed)
temperature field $T(\Vang)$ remapped by
the deflection vector,
\beeq
\lenT(\Vang)=T\left[\Vang+\ang\phi(\Vang)\right].
\label{eq:lensing}
\eneq
We will use notation with (or without) tilde to represent lensed (or unlensed)
quantities.
Note that we mainly work in the Rayleigh-Jeans tail and express the surface 
brightness in terms of temperature.

In a sufficiently small patch of the sky, it significantly simplifies the
manipulations to work in Fourier space
\citep[see][for all-sky formalism]{HU00,OKHU03,CHLE05}. In Fourier space
the lensed temperature is 
\bear
\label{eq:lensexp}
\lenT_\vL&=&\int d^2\Vang~\lenT(\Vang)~e^{-i\vL\cdot\Vang} \\ 
&=&T_\vL-\int{d^2\vL'\over(2\pi)^2}\left[(\vL-\vL')\cdot\vL'\right]T_{\vL'}
\phi_{\vL-\vL'}+\cdots, \nonumber
\enar
where we Taylor expanded $\lenT_\vL$ to the first order in $\phi_\vL$.
We kept the same notation for Fourier components, while the functional
dependence is indicated as a subscript (e.g., $T(\Vang)$ and $T_\vL$ are
Fourier counterparts).
The rms deflection angle $\langle\bdv{d}\cdot\bdv{d}\rangle^{1/2}$
is a few arcminutes and the deflection power peaks at a few degree scale,
comparable to the angular sizes of clusters.
However, the large-scale deflection field is coherent over the
scales of the temperature fluctuations, resulting in an unobservable overall
shift of the temperature field \citep{MAZA06}, and the linear approximation
remains valid. In Sec.~\ref{sec:mle} we discuss the limitation of this 
approximation when the lensing effect is large in a region around massive
clusters.

Since the intrinsic CMB is Gaussian and isotropic,
the statistical properties
of the temperature field can be completely described by the power spectrum
$C_l$,
\beeq
\langle T_{\vL_1}T^*_{\vL_2}\rangle=(2\pi)^2~\delta(\vL_1-\vL_2)~C_{l_1},
\eneq
where the asterisk represents complex conjugation and $\delta$ is the
Dirac delta function.
Analogously, we define the projected potential power spectrum $\Cphi_l$. 
Thus the deflection and the convergence
power spectra are $\Cdd_l=l^2\Cphi_l$ and $\Ckap_l=l^4\Cphi_l/4$, respectively.
Note that $\Cphi_l$ can always be defined in this way, though it may be an
incomplete description of the statistical properties of the projected potential
when $\phi_\vL$ is non-Gaussian. 
Finally, the power spectrum of the lensed temperature field is 
\beeq
\tilde C_l=\left[1-l^2R\right]C_l+\int{d^2\vL'\over(2\pi)^2}
\left[(\vL-\vL')\cdot\vL'\right]^2C_{l-l'}\Cphi_{l'},
\eneq
where $R\equiv(1/4\pi)\int d\ln l~l^4\Cphi_l$ 
is the half of the rms deflection angle \citep{HU00,LECH06}.

In practice, the observed temperature field has two additional contributions:
detector noise independent of the signal, and telescope beam
convolving the signals from different patches of the sky.
We assume that the detector noise is white, so that the noise
power spectrum is constant,
\beeq
C_l^N\equiv\Delta_T^2=\spix^2{4\pi \fsky\over \Npix},
\eneq
where $\spix$ is the rms error in each pixel of the detector 
in units of $\mu$K, 
$\fsky$ is the fraction of the survey area on the sky, and $\Npix$ is the
total number of detector pixels \citep{KNOX95}.
Convolution is simply a multiplication in Fourier space, and the
beam factor for a simple Gaussian beam we consider is
$B_l=\exp\left[-{1\over2}l^2\sbeam^2\right]$.
The beam width $\sbeam$ is related to the full-width half-maximum (FWHM) as
$\sbeam=\tfwhm/\sqrt{8\ln2}$.
The observed temperature field and its power spectrum are then
\bear
\label{eq:tobs}
\Tobs_\vL&=&\lenT_\vL~e^{-{1\over2}l^2\sbeam^2}+T^N_\vL,\\
\Cobs_l&=&\tilde C_l~e^{-l^2\sbeam^2}+C^N_l.
\enar
In reality, one needs to consider other contributions to $\Tobs$, such as
residual foregrounds, point radio sources, and 
CMB secondary anisotropies. We will
only consider secondary contributions in Sec.~\ref{ssec:sz}.

\subsection{Quadratic Estimator}
\label{ssec:qe}
Here we consider a convergence estimator $\khat(\Vang)$ that is quadratic
in the observed temperature field, accounting for  telescope beam and
detector noise.\footnote{We will use quantities with hat to represent
estimators of the quantities without hat, e.g., a convergence estimator
is denoted as $\khat$ and a true convergence field is denoted as $\kappa$.
However, this notational convention should not be confused with that
used for temperature fields: $T$, $\lenT$, $\Tobs$, and $\That$ represent
the intrinsic (unlensed), the lensed [Eq.~(\ref{eq:lensing})], the
observed [Eq.~(\ref{eq:tobs})], and the delensed [Eq.~(\ref{eq:delensT})]
temperature fields, respectively.}
We require that the estimator be unbiased when averaged
over an ensemble of CMB maps, $\langle\khat(\Vang)\rangle=\kappa(\Vang)$.
With these conditions, the estimator takes the general form in Fourier space
\beeq
\khat_\bdv{L}={N_L\over2}\int{d^2\vL_1\over(2\pi)^2}~F(\vL_1,\vL_2)~
\Tobs_{\vL_1}~\Tobs_{\vL_2},
\label{eq:khat}
\eneq
where $\vL_2=\bdv{L-l}_1$ and $N_L$ is a normalization coefficient, which
only depends on $L=|\bdv{L}|$.
The functional form of $F(\vL_1,\vL_2)$ can be obtained by minimizing the
variance of $\khat_\bdv{L}$ and imposing the normalization condition 
\beeq
F(\vL_1,\vL_2)={\left[\bdv{L}\cdot\vL_1C_{l_1}+
\bdv{L}\cdot\vL_2C_{l_2}\right]\over2~\Cobs_{l_1}\Cobs_{l_2}}~
e^{-{1\over2}l_1^2\sbeam^2}~e^{-{1\over2}l_2^2\sbeam^2},
\label{eq:fctform}
\eneq
and the normalization coefficient is
\beeq
{1\over N_L}={1\over L^2}\int{d^2\vL_1\over(2\pi)^2}
{\left[\bdv{L}\cdot\vL_1C_{l_1}+\bdv{L}\cdot\vL_2C_{l_2}\right]^2
\over2~\Cobs_{l_1}\Cobs_{l_2}}~e^{-l_1^2\sbeam^2}~e^{-l_2^2\sbeam^2}.
\eneq
Finally, the variance of the estimator is
\beeq
\langle\khat_\bdv{L}\khat^*_\bdv{L'}\rangle=(2\pi)^2~\delta(\bdv{L-L'})
(\Ckap_L+\Nkap_L),
\eneq
where $\Nkap_L=L^2N_L/4$ is the noise power spectrum of $\khat_\bdv{L}$.
One can think of 
$\Ckap_L/N_L^{\kappa\kappa}$ as a signal-to-noise ratio, and the reconstruction
becomes difficult at the angular scale $L$, where
$\Ckap_L\simeq N_L^{\kappa\kappa}$.
Given experimental specifications, the noise power spectrum 
$\Nkap_L$, as a function of the intrinsic CMB power spectrum $C_L$,
becomes smallest, when there exists substantial power in
$C_L$ at the scale of interest,
with its shape deviating from the scale-invariance ($L^2C_L=$constant)
\citep{ZAZA06}.

Our estimator recovers the general form of the standard quadratic estimators
as $\sbeam\rightarrow0$, and
$N_L$ corresponds to the noise power spectrum of a deflection
estimator $\hat\bdv{d}_\bdv{L}=2\bdv{L}~\khat_\bdv{L}/L^2$ 
used in the literature
\citep{HU01b}.

The estimator can be decomposed as two Wiener-filtered temperature functions
in real space, which essentially correlates the gradient of the lensed
temperature field with the unlensed temperature field to isolate the lensing
effect,
\bear
\label{eq:G}
\bdv{G}(\Vang)&=&\int{d^2\vL\over(2\pi)^2}~i\vL~\Tobs_\vL{C_l\over\Cobs_l}~
e^{-{1\over2}l^2\sbeam^2+i\vL\cdot\Vang} \\
\label{eq:W}
W(\Vang)&=&\int{d^2\vL\over(2\pi)^2}~\Tobs_\vL{1\over\Cobs_l}~
e^{-{1\over2}l^2\sbeam^2+i\vL\cdot\Vang},
\enar
and the convergence estimator can be expressed in terms of 
$\bdv{G}(\Vang)$ and $W(\Vang)$ as
\beeq
\khat_\bdv{L}=-{N_L\over2}i\bdv{L}\cdot\int d^2\Vang~\bdv{G}(\Vang)W(\Vang)
~e^{-i\bdv{L}\cdot\Vang}.
\label{eq:GW}
\eneq
This approach of using the two Wiener-filtered functions is more convenient 
for computing $\khat_\bdv{L}$ by using Fast Fourier Transform (FFT) routines 
than by directly computing Eq.~(\ref{eq:khat}). Furthermore, it is more 
physically intuitive than the general derivation, though the latter has clear
advantage in its transparency and understanding
the uniqueness of the functional form $F(\vL_1,\vL_2)$.
A modified quadratic estimator can be constructed by removing the signals
in Eq.~(\ref{eq:G}) at $l\geq\lcut$, while Eq.~(\ref{eq:W}) remains unchanged.

\begin{figure}
\centerline{\psfig{file=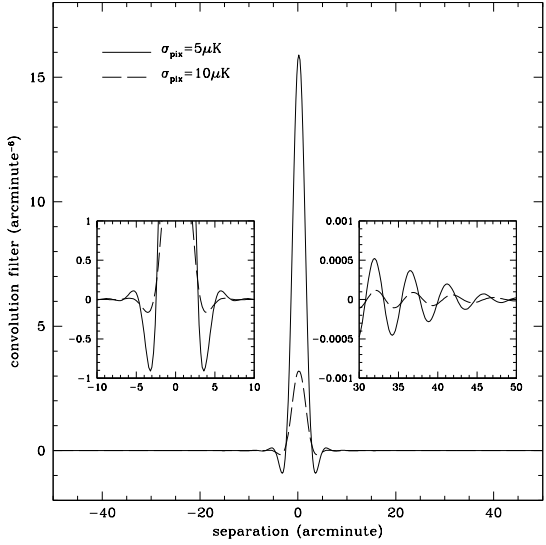, width=3.0in}}
\caption{Convolution filter $H(\theta)$ as a function of separation 
$\theta=|\Vang|$ for CMB experiments
with $\spix=5\muK$ and~$10\muK$ in Sec.~\ref{sec:num}.
The insets show details of $H(\theta)$ at the center ($left$)
and its tail ($right$).}
\label{fig:filter}
\end{figure}

To better understand how quadratic estimators operate, we Fourier transform and
rearrange Eq.~(\ref{eq:GW}) as
\bear
\label{eq:div}
{1\over2}\ang\cdot\left[\bdv{G}(\Vang)W(\Vang)\right]&=&
\int{d^2\bdv{L}\over(2\pi)^2}{-\khat_\bdv{L}\over N_L}~
e^{i\bdv{L}\cdot\Vang} \\
&=&\int d^2\Vangm~H(\Vangm-\Vang)\khat(\Vangm). \nonumber
\enar
The divergence of the two Wiener-Filtered functions is a convolution of the
convergence estimate $\khat(\Vang)$ and the filter
\beeq
H(\Vang)=\int{d^2\bdv{L}\over(2\pi)^2}~{-1\over N_L}~e^{i\bdv{L}\cdot\Vang}.
\label{eq:filterH}
\eneq
Figure~\ref{fig:filter} plots the filter $H(\theta)$ as a function of
separation $\theta=|\Vang|$ for experiments with $\spix=5\muK$ and~$10\muK$,
to which we apply quadratic estimators in Sec.~\ref{sec:num}. The filter
peaks at the center and its width is $\simeq3'$, roughly set by the scale 
that the intrinsic CMB and detector noise power spectra become comparable.
While the filter is highly oscillating at its tail, it is negligible
at $\theta\geq10'$ due to the large weight near the center.
A factor of two change in $\spix$ has little impact on the width of the 
filter, because the crossing scale is already at the CMB damping tail.

\section{Maximum Likelihood Estimator}
\label{sec:mle}
In this section, we analyze the likelihood of CMB lensing by singular
isothermal clusters. We first derive a quadratic estimator for singular 
isothermal clusters and compare the estimator to the optimal estimator
from the likelihood. With the simple singular isothermal model, 
our analysis will
be carried out analytically, showing that
(1)~the standard quadratic estimators are
unbiased and optimal in the limit of no lensing, (2)~they progressively become
biased and sub-optimal when the lensing effect increases, 
and (3)~why the modified quadratic
estimators perform better than the standard quadratic estimators. Finally, we
develop a unbiased maximum likelihood estimator to reconstruct 
cluster mass profiles as well as large-scale structure.
We demonstrate its applicability to CMB experiments with tests
against numerical simulations using more realistic cluster models
in Sec.~\ref{sec:num}.

\subsection{Quadratic Estimator for a Singular Isothermal Cluster}
A singular isothermal cluster has a density profile $\rho(r)\propto r^{-2}$
and its enclosed mass increases with $r$, which requires truncation at some
radius to be a viable model for real clusters. However, this
model has advantage in its simplicity: its 
properties are described by one parameter, Einstein radius

\beeq
\tE=4\pi\sigma^2~{D_\star-D_L\over D_\star},
\eneq
where $\sigma$ is one-dimensional velocity dispersion of a cluster
and $D_L$ is
the comoving angular diameter distance to the lensing cluster. CMB
lensing has a well-defined single plane of the source redshift and
the comoving angular diameter distance to the last scattering surface 
$D_\star=14.12$~Gpc is now measured with less than 1\% uncertainty 
\citep{KODUET08}.
The convergence is $\kappa(\Vang)=\tE/2\theta$ and the deflection 
vector is $\bdv{d}(\Vang)=-\tE\Vang$ given the angular separation 
$\theta=|\Vang|$ from the origin
in a cluster centric coordinate. When a virial radius $\Rvir$
is defined as the radius inside which the mean density is 200 times the 
cosmic mean matter density,
a singular isothermal cluster of mass $M=10^{14}\hmsun$ 
within the virial
radius at $z_L=1$ has an Einstein radius $\tE=8.\!\!''0$ and a velocity
dispersion $\sigma=2.0\times10^{-3}(=610~\kms)$, 
and they scale as $\tE\propto M^{2/3}$ and $\sigma\propto M^{1/3}$.

A quadratic estimator $\tEQE$
for singular isothermal clusters can be readily derived
using the method described in Sec.~\ref{ssec:qe}, but here we take an idealized
approach for the purpose of comparison, where we assume $\spix=\sbeam=0$.
Under the condition
that the estimator is unbiased $\langle\tEQE\rangle=\tE$ and it has the
minimum variance, the quadratic estimator is
\bear
\label{eq:teqe}
\tEQE&=&{1\over\Fisher}\int{d^2\vL_1\over(2\pi)^2}\int{d^2\vL_2\over(2\pi)^2}
\\
&\times&{\lenT_{\vL_1}\lenT_{\vL_2}\over \lenC_{l_1}\lenC_{l_2}}
{\pi(\vL_1C_{l_1}+\vL_2C_{l_2})\cdot(\vL_1+\vL_2)\over|\vL_1+\vL_2|^3},
\nonumber
\enar
with the normalization coefficient 
\bear
\label{eq:fisher}
\Fisher&=&\int{d^2\vL_1\over(2\pi)^2}\int{d^2\vL_2\over(2\pi)^2}\\
&\times&{2\pi^2\over \lenC_{l_1}\lenC_{l_2}} 
\left[{(\vL_1C_{l_1}+\vL_2C_{l_2})\cdot(\vL_1+\vL_2)
\over|\vL_1+\vL_2|^3}\right]^2. \nonumber
\enar
The variance of the estimator is 
$\langle(\tEQE-\tE)(\tEQE-\tE)\rangle=1/\Fisher$.
Here we Taylor expanded $\lenT_\vL$ and kept terms only to the first order
in $\tE$ in deriving $\tEQE$.

\subsection{Relation to the Optimal Estimator}
\label{ssec:optimal}
The likelihood function $P(\lenT|\tEm)$ simply represents the probability that
a singular isothermal model 
with $\tEm$ can have the lensed temperature field $\lenT(\Vang)$. Since the
intrinsic CMB follows a Gaussian distribution and gravitational lensing only
remaps the intrinsic CMB, the distribution of $\lenT(\Vang)$ is also
Gaussian and its statistical properties are fully described by the covariance
matrix of $\lenT(\Vang)$
\beeq
\lenC(\Vang,\Vang')=\langle\lenT(\Vang)\lenT(\Vang')\rangle
=\int{d^2\vL\over(2\pi)^2}~\lenC_l ~e^{i\vL\cdot(\Vang-\Vang')}.
\eneq
For convenience, we take a negative logarithm of $P(\lenT|\tEm)$ and call it
likelihood,
\bear
\label{eq:likeli}
\Lik(\lenT|\tEm)&\equiv&-\ln P(\lenT|\tEm) \\
&=&{1\over2}\lenT(\Vang)~\lenC^{-1}(\Vang,\Vang'|\tEm)~\lenT(\Vang')+
{1\over2}\ln\det\lenC(\tEm), \nonumber
\enar
where the summation over $\Vang$ and $\Vang'$ is implicitly assumed and 
hereafter we will suppress the angular dependence for simplicity.
In general, the likelihood is a functional with its argument of a scalar field,
such as $\kappa(\Vang)$ or $\phi(\Vang)$. However, in our case
it reduces to a function 
with its argument of a scalar $\tEm$, substantially simplifying
the manipulation.

We take a derivative of $\Lik$ with respect to $\tEm$,
\bear
\label{eq:first}
{\partial\Lik\over\partial\tEm}&=&-{1\over2}\lenT~\lenC^{-1}~
{\partial\lenC\over\partial\tEm}~\lenC^{-1}~\lenT \\
&=&-\int\!\!\!{d^2\vL_1\over(2\pi)^2}\int\!\!\!{d^2\vL_2\over(2\pi)^2}
{\lenT_{\vL_1}\lenT_{\vL_2}\over\lenC_{l_1}\lenC_{l_2}}
{\pi(\vL_1C_{l_1}+\vL_2C_{l_2})\cdot(\vL_1+\vL_2)\over|\vL_1+\vL_2|^3}, \nonumber
\enar
where we computed the derivative to the first order in $\tEm$.
Since gravitational lensing only redistributes the intrinsic CMB,
the last term (log determinant) in Eq.~(\ref{eq:likeli}) is 
independent of $\tEm$ and
hence the derivative with respect to $\tEm$ vanishes in Eq.~(\ref{eq:first}).
However, in the presence of non-white instrumental noise, 
and/or other secondary
contaminants, the derivative acquires a nonzero value but it is in general 
negligible compared to the quadratic term in Eq.~(\ref{eq:first}). We will
neglect this effect in the remainder of this paper.
In the presence of significant contaminants
from secondaries, the assumption that the likelihood function is Gaussian
becomes invalid before the log determinant term becomes non-negligible.

With the derivative of $\Lik$, we can compute the Fisher information matrix
\beeq
\label{eq:fisher2}
\Fisher=\left\langle{\partial^2\Lik\over\partial\tEms}\right\rangle
=\left\langle{\partial\Lik\over\partial\tEm}{\partial\Lik\over\partial\tEm}
\right\rangle 
\eneq
where for the second equality we used the normalization condition of the
likelihood function
$1=\int d\lenT~ P(\lenT|\tEm)=\int d\lenT~ e^{-\Lik}$. Within the Gaussian
approximation, $\Fisher$ can be evaluated at any value of $\tEm$.
Note that $\Fisher$ is identical to the normalization coefficient in 
Eq.~(\ref{eq:fisher}). 

In statistical parameter estimation, there exists a powerful theorem, known
as the Cram\'er-Rao inequality that error bars in a parameter estimation
have a definite lower bound  $\sigma(\tEm)\geq\Fisher^{-1/2}$
set by the Fisher matrix.
Moreover, this theorem provides a necessary and sufficient condition for an
estimator to saturate the Cram\'er-Rao inequality, i.e., to be an optimal 
estimator $\tEopt$ \citep{BABIC05},
\beeq
{\partial\Lik\over\partial\tEm}=\Fisher~(\tEm-\tEopt).
\label{eq:optimality}
\eneq
Now it is apparent that only in the limit of no lensing 
(the true Einstein radius $\tE=\tEm=0$) does the
quadratic estimator $\tEQE$ become an optimal estimator $\tEopt$ with the
smallest variance attainable from the data. Conversely, $\tEQE$ becomes
progressively biased and sub-optimal as the lensing effect increases.
This can be also understood by the validity of the linear approximation:
since the quadratic estimator is constructed to be unbiased and to minimize
the variance when $\lenT_\vL$ is expanded to the linear order in $\phi_\vL$, 
it is natural to expect that this condition breaks down when higher-order terms
in $\phi_\vL$ become dominant over the linear order term. The modified 
quadratic estimator, on the other hand, removes the angular modes of the
signals at $l\geq\lcut$ by explicitly setting the integrand zero
in Eq.~(\ref{eq:teqe}), where the linear approximation breaks down, and this 
process helps suppress the contributions from the higher-order terms 
in $\phi_\vL$
because the higher-order terms are related to multiple integrals over the modes
that are suppressed most. Precisely for this reason could the modified
quadratic estimators be more robust than the standard quadratic estimators
even when the lensing effect is large.

However, the modified quadratic estimator requires a rather arbitrary choice 
of the cutoff scale $\lcut$, which depends on the lensing effect,
though it may be possible to calibrate against simulations \citep{HUDEVA07}. 
Furthermore, the removal of the lensing signals at $l\geq\lcut$ inevitably
results in lower signal-to-noise ratio, making the reconstruction noisier.
We discuss this issue with numerical simulations in Sec.~\ref{ssec:com}.

\subsection{Maximum Likelihood Estimator}
Given the Gaussian probability distribution of the CMB, 
the likelihood retains all the information of the observed data.
Even when there exists no optimal estimator, one can always find an estimator,
if not analytically, that maximizes the likelihood:
the maximum likelihood estimator $\tEML$ is the solution of
\beeq
{\partial\Lik\over\partial\tEm}\Bigg|_{\tEm=\tEML}=0.
\label{eq:mle}
\eneq
However, this equation is highly non-linear in general and requires 
approximations to be solved even numerically. 
Equations~(\ref{eq:optimality}) and~(\ref{eq:mle}) show that an optimal 
estimator is always the maximum likelihood estimator. However, note that
while the converse is not true in general, the 
maximum likelihood estimator asymptotically 
approaches to the optimal condition. 

Having understood that the quadratic estimator becomes an optimal (and maximum
likelihood) estimator in the limit of no lensing in Sec.~\ref{ssec:optimal},
we present an alternative approach to modeling the likelihood and derive a
new maximum likelihood estimator for singular isothermal clusters. We then 
generalize this approach to clusters with arbitrary mass distributions.

Consider a model with $\tEm$ and its deflection field 
$\bdv{d}^m(\Vang)=-\tEm\Vang$.
We construct a delensed
temperature field $\That(\Vang)$ by delensing the observed
$\lenT(\Vang)$ with $\bdv{d}^m(\Vang)$, and $\That(\Vang)$ is related to
the intrinsic temperature field $T(\Vang)$ as
\bear
\label{eq:delensT}
\That(\Vang)&\equiv&\lenT(\Vang-\bdv{d}^m) \\
&=&T(\Vang-\bdv{d}^m+\bdv{d})=T\left[(1+\Delta)\Vang\right], \nonumber
\enar
with $\Delta=\tEm-\tE$. Now we can write the likelihood in terms of the 
delensed temperature field $\That(\Vang)$
\beeq
\Lik(\That|\tEm)={1\over2}\That(\tEm)~C^{-1}~\That(\tEm)+{1\over2}\ln\det C,
\eneq
where we emphasized the dependence of $\That(\Vang)$ on $\tEm$,
and $C$ is the covariance matrix of $T(\Vang)$.
Taking a derivative of $\Lik$ with respect to $\tEm$ gives
\bear
{\partial\Lik\over\partial\tEm}&=&{1\over2}\left[{\partial\That\over\partial
\tEm}C^{-1}~\That+\That~ C^{-1}{\partial\That\over\partial\tEm}\right] \\
&=&-\int\!\!\!{d^2\vL_1\over(2\pi)^2}\int\!\!\!{d^2\vL_2\over(2\pi)^2}
{T_{\vL_1}T_{\vL_2}\over C_{l_1}C_{l_2}}
{\pi(\vL_1C_{l_1}+\vL_2C_{l_2})\cdot(\vL_1+\vL_2)\over|\vL_1+\vL_2|^3}. 
\nonumber
\enar
The second equality is obtained by evaluating the derivative at $\Delta=0$.
Assuming that our initial
model with $\tEst$ is a good approximation to the true model with $\tE$
($\Delta_\star=\tEst-\tE\simeq0$), the 
likelihood can be expanded around $\Delta_\star$
\bear
\Lik&=&\Lik_\star+\left({\partial\Lik\over\partial\tEm}\right)_\star
(\Delta-\Delta_\star) \\
&+&{1\over2}\left({\partial^2\Lik\over\partial\tEms}
\right)_\star(\Delta-\Delta_\star)^2+{\mathcal O}(\Delta^3),\nonumber
\enar
and we can use the standard Newton-Raphson method to solve Eq.~(\ref{eq:mle})
and obtain a maximum likelihood estimator $\tEML$,
\bear
\label{eq:iterD}
\Delta(\tEML)-\Delta_\star&=&\tEML-\tEst \\
&=&-\left({\partial\Lik\over\partial\tEm}\right)_\star\bigg/
\left({\partial^2\Lik\over\partial^2\tEms}\right)_\star.\nonumber
\enar
It is important to note that the validity of our solution for $\tEML$
is independent of the linear approximation, but
the convergence of $\tEML$ depends on the goodness of $\tEst$ 
to $\tE$. Eq.~(\ref{eq:iterD}) still involves computationally
intensive evaluations of the second derivative, or the curvature matrix.
We further simplify $\tEML$ by replacing the curvature matrix with its 
ensemble average, Fisher matrix
\bear
\hat\Fisher&=&\int{d^2\vL_1\over(2\pi)^2}\int{d^2\vL_2\over(2\pi)^2} \\
&\times&{2\pi^2\over C_{l_1}C_{l_2}}
\left[{(\vL_1C_{l_1}+\vL_2C_{l_2})\cdot(\vL_1+\vL_2)
\over|\vL_1+\vL_2|^3}\right]^2, \nonumber
\enar
and by evaluating the derivatives at $\Delta_\star=0$.
Finally, our new maximum likelihood estimator is
\bear
\label{eq:iter}
\tEML&=&\tEst+{1\over\hat\Fisher}\int{d^2\vL_1\over(2\pi)^2}
\int{d^2\vL_2\over(2\pi)^2} \\
&\times&
{\That_{\vL_1}\That_{\vL_2}\over C_{l_1}C_{l_2}}
{\pi(\vL_1C_{l_1}+\vL_2C_{l_2})\cdot(\vL_1+\vL_2)\over|\vL_1+\vL_2|^3}. \nonumber
\enar
This equation is readily recognizable as the standard quadratic estimator
in Eq.~(\ref{eq:teqe}), except $\lenC_l$ and $\lenT_\vL$ replaced with $C_l$ and
$\That_\vL$. The resemblance should not be surprising, and in hindsight one
could have expected this outcome
given the result in Sec.~\ref{ssec:optimal}: the quadratic
estimator becomes optimal when the lensing effect is vanishingly small; 
as we delens $\lenT(\Vang)$ well enough that $\That(\Vang)$ is close to 
$T(\Vang)$, the residual lensing effect in $\That(\Vang)$ is substantially 
reduced and therefore the maximum likelihood estimator takes
the form of the quadratic estimator, returning diminishing change of the
second term in Eq.~(\ref{eq:iter}), i.e., 
$\tEML\simeq\tEst\simeq\tE$.

We want to emphasize that this new estimator in the form of quadratic 
estimators is derived by iteratively solving for the maximum likelihood
in Eq.~(\ref{eq:mle}) and updating the initial model $\tEst$ as in the
standard Newton-Raphson method,
i.e., it is a maximum likelihood estimator and is
independent of the linear approximation, to which the validity of the standard
quadratic estimator is limited. One may be concerned about replacing the
curvature matrix with the Fisher matrix in Eq.~(\ref{eq:iter}) and obtaining
a solution quadratic in $\That_\vL$ instead of a solution rational in 
$\That_\vL$ (quadratic in $\That_\vL$ both in numerator and in denominator).
However, both procedures guarantee that the correct solution of 
Eq.~(\ref{eq:mle}) is iteratively found reaching the same peak of the 
likelihood, while the error estimation of parameters is approximated by using
the Fisher matrix, rather than the full curvature matrix. In Sec.~\ref{sec:num}
we demonstrate that this is a good approximation and the initial
model converges quickly to the true model.
Given the nomenclature of the existing quadratic estimators, now let us call
our new maximum likelihood estimator an improved quadratic 
estimator.\footnote{However, note 
that since our new estimator takes the result
of the previous iteration as an initial model, another iteration makes use 
of $\That(\Vang)$ that is constructed by using the initial model and this
initial model is also a function of $\That(\Vang)$ in the previous iteration,
which makes the estimator a rational function of temperature, instead of a
quadratic function. Therefore,  it is technically incorrect to call it 
a quadratic estimator.}

In practice we can use the standard quadratic
estimators to obtain an initial model and 
then proceed with our improved quadratic
estimator to refine the solution, even when the lensing effect is large. 
In general, the reconstruction of cluster mass profiles
is too noisy to provide a good initial model.
However, we can adopt an initial model for clusters from other 
observations (e.g., galaxy weak lensing and X-ray measurement) or theoretical 
expectations (e.g., Navarro-Frenk-White (NFW) profiles \citep{NFW97}). As 
opposed to the modified quadratic estimators, there is no arbitrary choice of
$\lcut$ in our method. 

The toy model developed here can be readily generalized and our
improved quadratic estimator can be used to reconstruct mass profiles of
realistic clusters and large-scale structure. However, in the presence of
the telescope beam and detector noise, the delensing process becomes 
non-optimal because it does not commute with the beam smoothing. In the
absence of detector noise, one can deconvolve the beam factor, delens the
temperature field, and convolve the beam again, which can solve the problem
of non-commutativity. 

However, in the presence of detector noise, the beam
deconvolved noise can produce unwanted power on all scales when it is
delensed due to the non-white power below the beam scale. One can in principle
filter out or remove these small scales before delensing to mitigate the 
problem \citep{HUDEVA07}, 
which however introduces additional ad hoc scale to the problem.
The impact of telescope beam and detector noise is small in practice
for surveys like SPT ($\Delta_T\simeq6\muK$-arcmin) and ACT 
($\Delta_T\simeq10\muK$-arcmin)
as we numerically demonstrate in 
Sec.~\ref{sec:num}. We explicitly show in Appendix~\ref{app:delens}
that the delensing process suppresses the beam effect by a factor of the
average magnification by clusters,
since it corresponds to a mapping from the image plane to the source plane.
Non-white instrumental noise and boundary effect of
detectors may affect the delensing process. However, compared to the survey
area, the lensing signals are limited to a relatively small region 
around clusters where none of those effect is expected to be significant.

\section{Reconstructing Cluster Mass Profiles}
\label{sec:num}
Here we use numerical simulations of the CMB and cluster lensing potential to 
demonstrate the applicability of our improved quadratic estimator to 
CMB experiments. First, we adopt a more realistic model for massive 
clusters and investigate the dependence of our improved
quadratic estimator on assumed initial models in Sec.~\ref{ssec:iQE}. Then we 
reconstruct cluster mass profiles using the standard, modified, and
improved quadratic estimators, and we compare their performance 
in Sec.~\ref{ssec:com}.
Finally, we discuss the effects of contaminants and investigate the robustness
of our improved quadratic estimators in the presence of the Sunyaev-Zel'dovich
(SZ) effects.

\begin{figure}
\centerline{\psfig{file=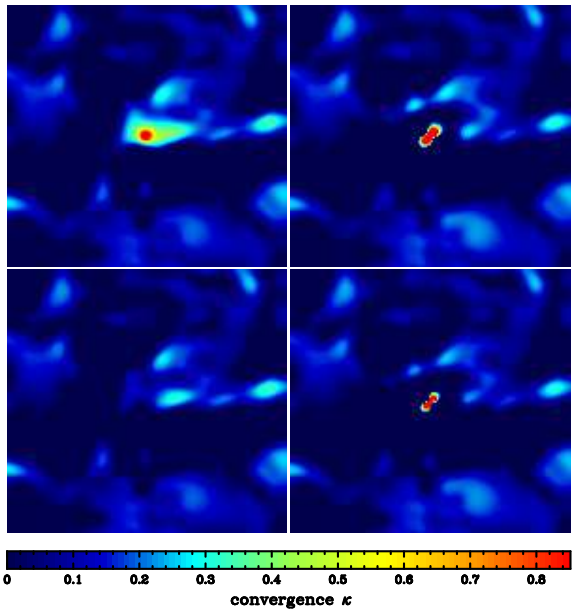, width=3.0in}}
\caption{(color online)
Reconstructed convergence fields of a $30'\times30'$ region around
a cluster at $z_L=1$ from an ideal experiment with $\Delta_T=0$. 
Cluster mass is set $M=5\times10^{14}\hmsun$.
Improved quadratic estimators are applied once with initial mass models of 
$\Minit=5\times10^{14}\hmsun$ ({\it left})
and $\Minit=1\times10^{14}\hmsun$ ({\it right}) to a single patch of sky. 
The bottom
panels show the residual after the true cluster convergence field is 
subtracted from the top panels.}
\label{fig:residual}
\end{figure}

\subsection{Improved Quadratic Estimator}
\label{ssec:iQE}
A singular isothermal model used in Sec.~\ref{sec:mle} is useful in developing
an analytic solution of the likelihood approach. However, it is rather an
academic model than a realistic model for massive clusters.
Recent numerical simulations show that there exist a universal mass profile
for dark matter halos, NFW profiles \citep{NFW97}
\beeq
\rho(r)={\rho_s\over r/r_s(1+r/r_s)^2}.
\eneq
The scale radius $r_s$ is described by the concentration parameter 
$c=\Rvir/r_s$ and the normalization coefficient $\rho_s$ 
is related to the mass of clusters $M=4\pi r_s^3\rho_s[\ln(1+c)-c/(1+c)]$. 
We now use NFW profiles to model massive clusters.

The convergence field $\kappa(\Vang)$
of NFW profiles can be obtained by the ratio of 
the projected mass density $\Sigma(r)$ to the critical surface density 
$\Sigma_\up{crit}$ of the lensing cluster at $z_L$,
\beeq
\kappa\left(\theta={r\over D_L}\right)={\Sigma(r)\over\Sigma_\up{crit}}=
{2~r_s\rho_s\over\Sigma_\up{crit}}
P\left({r\over r_s}\right)(1+z_L)^2,
\label{eq:kappa}
\eneq
where the functional form $P(x)$ of the projected density is 
\citep{BARTE96,WRBR00}
\bear
P(x)&=&{1\over x^2-1}\left[1-{2\over\sqrt{1-x^2}}~\up{tanh}^{-1}
\sqrt{1-x\over1+x}\right],~~~(x<1) \nonumber \\
&=&{1\over3},~~~(x=1)\\
&=&{1\over x^2-1}\left[1-{2\over\sqrt{x^2-1}}~\up{tan}^{-1}\sqrt{x-1\over x+1}
\right],~~~(x>1)\nonumber
\enar
and the critical surface density 
$\Sigma_\up{crit}^{-1}=4\pi G D_L(D_\star-D_L)/D_\star(1+z_L)$ is only a 
function
of $z_L$ given the precise measurement of $D_\star$. Note that the convergence
field $\kappa$ of NFW profiles depend only on the angular separation
$\theta=|\Vang|$ due to spherical symmetry.
The redshift dependence in Eq.~(\ref{eq:kappa})
arises due to our use of comoving coordinates, reflecting higher densities 
of the universe at $z_L>0$. For reference, $D_L=850\hmpc$ and $2400\hmpc$, and
$\Sigma_\up{crit}=2.8\times10^3hM_\odot\up{pc}^{-2}$ 
and $1.8\times10^3hM_\odot\up{pc}^{-2}$ for
$z_L=0.3$ and~1, respectively. For clusters of $M=5\times10^{14}\hmsun$ and 
$1\times10^{14}\hmsun$, $\Rvir=2.1\hmpc$ and $1.2\hmpc$ appear subtended by
$3.\!'0$ and $4.\!'9$ on the sky at $z_L=1$ and 0.3.

We use {\scriptsize CMBFAST} \citep{SEZA96} to 
generate CMB temperature maps of $200'\times200'$ ($1000\times1000$ pixels)
and set the pixel scale $0.\!'2$ smaller than detector beam sizes. Given
a cluster mass $M$ and redshift $z_L$, we first compute the convergence field
$\kappa(\Vang)$ using Eq.~(\ref{eq:kappa}). The lensing potential $\phi(\Vang)$ 
and its deflection vector $\bdv{d}(\Vang)$ of the cluster are then computed
in Fourier space, where their relations to $\kappa(\Vang)$ become a simple
multiplication.
The lensed temperature field $\lenT(\Vang)$ is computed by displacing
the intrinsic temperature field $T(\Vang)$ with $\bdv{d}(\Vang)$ according to
Eq.~(\ref{eq:lensing}). Finally, we smooth $\lenT(\Vang)$ with a telescope
beam and add detector noises to obtain $\Tobs(\Vang)$.
Standard quadratic estimators can be used to reconstruct a convergence
field $\khat(\Vang)$ by using Eqs.~(\ref{eq:G}), (\ref{eq:W}), and 
(\ref{eq:GW}) with $\Tobs(\Vang)$, and so can modified quadratic estimators
with a choice of $\lcut$, beyond which the integrand in Eq.~(\ref{eq:G})
is set zero.

Similarly, our new estimation process begins with finding a solution
$\bhat$ to the delensing equation $\bhat=\Vang+\ang\phi^m(\Vang)$ given the
lensing potential $\phi^m(\Vang)$ of an assumed initial model. We then 
construct a delensed temperature field $\That(\bhat)=\Tobs(\Vang)$ and
use the same equations with $\Tobs(\Vang)$ replaced by $\That(\bhat)$
to reconstruct $\khat_\bdv{L}$. Imposing a consistency condition between
the assumed model and the estimation result can provide a criterion for the
iteration convergence of our improved quadratic estimators.

ACT and SPT will find $\sim2\times10^4$
massive clusters mainly by the spectral distortion of 
the CMB arising from the inverse Compton scattering of hot electrons in 
clusters, so called the SZ effect \citep{SUZE70,SUZE72},
with roughly redshift-independent threshold mass 
$M\geq2\times10^{14}\hmsun$. To test our improved quadratic estimators,
we consider a typical cluster of $M=5\times10^{14}\hmsun$ and $c=3$.
Figure~\ref{fig:residual} shows the reconstructed $\khat(\Vang)$ of a 
massive cluster at $z_L=1$ in an ideal experiment with $\Delta_T=0$. 
Here we simply adopt a NFW profile with fixed concentration
$c=3$ for our initial model 
and allow mass $\Minit$ of the model to vary. Even with fixed concentration,
$r_s$ changes as a function of $\Minit$, and hence our assumption allows
for changes in the shape as well as the scaling of initial mass models. 
However, note that while we use this parametrized model of clusters,
our reconstruction is general and non-parametric, such that we recover
2-D structure of
$\kappa(\Vang)$ at each pixel rather than obtain model parameters $M$ and $c$
(see \citep{DODEL04,LEKI06} for reconstructing a parametrized cluster model). 
We assume that the cluster center is known from other observations with 
uncertainty less than our pixel scale $0.\!'2$.
The upper panels show the reconstructed $\khat(\Vang)$ from our
improved quadratic estimator using an initial model of
$\Minit=5\times10^{14}\hmsun$ ({\it left}) and $1\times10^{14}\hmsun$ 
({\it right}), and the bottom panels
show the residual after the true $\kappa(\Vang)$ is subtracted from the top
panels.

\begin{figure}
\centerline{\psfig{file=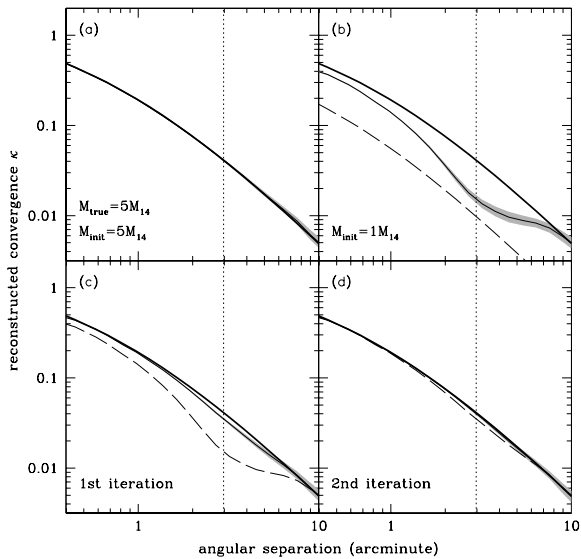, width=3.0in}}
\caption{Dependence of reconstructed mass profiles on an initial mass model
$\Minit$. Thick and thin solid lines represent the true cluster mass profile 
and the mean of reconstructed mass profiles from 500 clusters. 
The mass profiles are obtained by averaging reconstructed convergence
over the annulus of each cluster. 
The uncertainties in the mean profile are shown as shaded regions.
Dashed lines show an assumed
initial mass model and the cluster virial radius is shown as vertical
dotted lines. In Panels~($c$) and~($d$), the initial mass models are 
taken as the mean mass profile from the previous
iteration. The reconstruction quickly converges to the true mass profile 
in two iterations even with an incorrect 
choice of $\Minit=1\times10^{14}\hmsun$, 
exhibiting no detectable bias in an ideal experiment.}
\label{fig:iQE}
\end{figure}

With the perfect initial model in the left panels, the delensed temperature
field $\That(\Vang)$ is identical to the intrinsic $T(\Vang)$, and our improved
quadratic estimator returns $no$ change on average to the initial model
({\it bottom}). However, there exist random noises in $\khat(\Vang)$ over the
map, arising from the fluctuations of the intrinsic temperature gradient,
though they are evidently small and discernible from the massive cluster
({\it top}). In the right panels, $\lenT(\Vang)$ is delensed with the imperfect
initial model, so that $\That(\Vang)$ is not identical to $T(\Vang)$ but 
the lensing effect is significantly reduced.
In this regime, quadratic estimators become asymptotically
optimal and reconstruct $\kappa(\Vang)$ unbiased. The top panel exhibits
small anisotropy and some residual remains in the bottom panel. In a single
patchy of the sky, the CMB anisotropy has a gradient direction and 
gravitational lensing of the CMB makes no difference orthogonal to the 
gradient direction, in which reconstruction is completely degenerate,
resulting in the asymmetry in $\khat(\Vang)$.
However, since the CMB has no preferred direction, this obstacle can be 
overcome by stacking clusters in different patches of the sky. In practice,
this stacking process provides the average $\kappa(\Vang)$ of the clusters,
or the cluster-mass
cross-correlation function \citep{HUDEVA07}. Hereafter we assume that identical
clusters are stacked for simplicity.

We now quantify the ability to reconstruct $\kappa(\Vang)$ with varying
accuracy of assumed models. Figure~\ref{fig:iQE} plots the reconstructed
cluster mass profiles from 500 clusters ({\it thin solid}). The mass profiles
are obtained by averaging reconstructed $\khat(\Vang)$ over the annulus of 
each cluster, and the uncertainties in the mean mass profile
are shown as shaded 
regions. Figure~\ref{fig:iQE}$a$ shows that our improved quadratic
estimator is unbiased when our assumed model is perfect; it 
recovers the true model ({\it thick solid}) with no bias. If an 
assumed initial model is significantly different from the true model
in Fig.~\ref{fig:iQE}$b$,
the improved quadratic estimator suffers from the same problem
that the standard quadratic estimators have, and the reconstruction is again
biased low when the residual lensing effect is large. However, the 
reconstructed $\khat(\Vang)$ is inconsistent with our assumed model
({\it dashed}), implying that it has not converged to the correct solution.
In Fig.~\ref{fig:iQE}$c$ we take the reconstructed $\khat(\Vang)$ as a new
initial model and apply our improved quadratic estimator to the same clusters.
The reconstructed $\khat(\Vang)$ is now close to the true $\kappa(\Vang)$,
but still inconsistent with the assumed model. We iterate once more in 
Fig.~\ref{fig:iQE}$d$ and the reconstructed $\khat(\Vang)$ is identical to
the true $\kappa(\Vang)$. One more iteration results in no further change
and the estimate is consistent with the assumed and also the true models,
indicating the convergence of our estimates.

Even with the imperfect initial model, the
reconstruction quickly converges to the true $\kappa(\Vang)$ and 
no significant bias develops even beyond $\Rvir$ ({\it dotted}).
When the reconstructed $\khat(\Vang)$ is inconsistent with the assumed model,
one can in principle adopt a different initial model for a faster convergence
before applying the estimator iteratively. Note that the asymmetry seen
in Fig.~\ref{fig:residual} disappears and the reconstructed $\khat(\Vang)$
restores symmetry,
once many clusters are stacked. Furthermore, the uncertainties in the mean
profile decrease as our assumed model converges to the true model, because
it solely results from the intrinsic fluctuations of the CMB in the case of
perfect delensing.

\begin{figure}
\centerline{\psfig{file=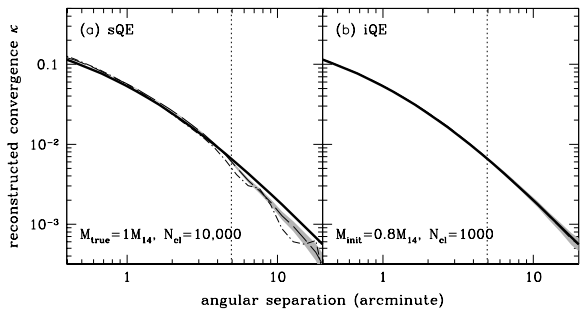, width=3.0in}}
\caption{Mass profile reconstruction for low mass clusters 
of $M=1~\times10^{14}\hmsun$ at $z_L=0.3$ from standard
(sQE) and improved (iQE) quadratic estimators (in the same format as in 
Fig.~\ref{fig:iQE}). 10,000 ({\it left}) and 1000 ({\it right}) clusters 
are used to obtain the mean profile, and the shaded region shows the 
uncertainties in the mean profile. Both estimators recover the true mass
profiles within $\Rvir$ in the low mass regime.
Approximately ten times more clusters are needed for sQE to achieve
the same accuracy than for iQE. However, for comparison we plot the mean 
profile from 1000 clusters as the dot-dashed line in the left panel.}
\label{fig:lowM}
\end{figure}

\begin{figure*}
\centerline{\psfig{file=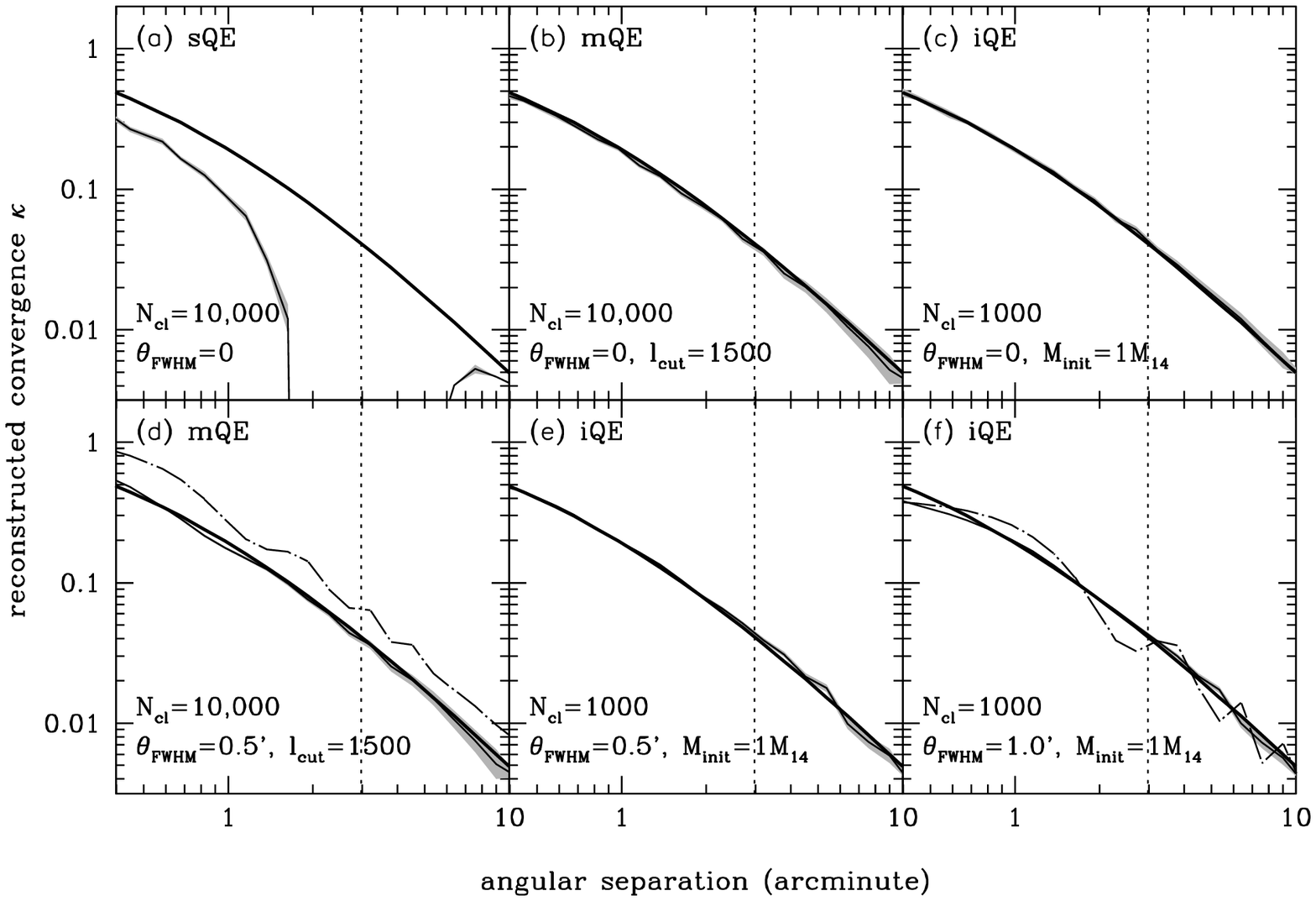, width=5.5in}}
\caption{Comparison of reconstructed mass profiles from standard (sQE), 
modified (mQE), and improved (iQE) quadratic estimators 
in realistic experiments with $\spix=5\muK$. 
The reconstruction is more difficult in the presence of detector noise and
telescope beam. For the mean of reconstructed mass profiles,
10,000 clusters of $M=5\times10^{14}\hmsun$ at $z_L=1$ are stacked when sQE or 
mQE is used, while iQE is iteratively applied to only 1000 clusters. 
The shaded regions show the uncertainties in the mean profile.
The dot-dashed line (panel~$d$) shows the shape distortion in $\khat(\Vang)$
when mQE is applied after beam-deconvolution, and the line is displaced to 
avoid confusion (see the text). With $\tfwhm=1'$ (panel~$f$),
iQE can recover the mean mass profile 
with small bias below the beam scale. For comparison, we plot
the reconstructed mass profile ({\it dot-dashed}) using mQE in Panel~($f$).}
\label{fig:com}
\end{figure*}

\subsection{Performance Comparison}
\label{ssec:com}
Before we assess the performance of the three lensing estimators 
in realistic experiments,
we first compare our improved quadratic estimator to the standard quadratic
estimator, when the lensing effect is small. Figure~\ref{fig:lowM} plots
the reconstructed cluster mass profiles in the same format as 
Fig.~\ref{fig:iQE}. For clusters of $M=1\times10^{14}\hmsun$ at $z_L=0.3$ 
($\kappa\ll1$),
the improved quadratic estimator recovers the true mass profile with no 
detectable bias after two iterations. With signals smaller by
a factor of five than in Fig.~\ref{fig:iQE}, 1000 clusters are
 stacked to obtain the mean mass profile, while 10,000 clusters are
required for the standard quadratic estimator. As we quantify the
difference in the signal-to-noise ratio below, 
the standard quadratic estimator needs approximately 
ten times as many clusters as the improved quadratic estimator needs to
achieve the same accuracy, 
but we show the mean profile ({\it dot-dashed}) obtained by applying the
standard quadratic estimator to 1000 clusters for comparison. Once enough
clusters are stacked, the standard quadratic estimator
works well within $\Rvir$, though it shows some hint of deviation
at the core. Thus, the standard quadratic
estimator may be safely used to reconstruct mass profiles of clusters with
$M<1\times10^{14}\hmsun$ at $z_L=0.3$. However, given the source of the CMB
at $z_\star=1090$, the lensing effect becomes larger
as $z_L$ increases, until $\Sigma_\up{crit}$ reaches the minimum at 
$z_L\simeq2.5$, where $D_L$ becomes a half of $D_\star$. Therefore, the standard
quadratic estimator cannot be used to reconstruct unbiased mass profiles of
clusters that are either at $z_L\geq0.3$ or massive 
$M\geq1\times10^{14}\hmsun$. Since
ACT and SPT will find clusters of $M\geq2\times10^{14}\hmsun$ 
at higher redshift, modified
or improved quadratic estimators are preferred to the standard quadratic 
estimator.

Now we consider realistic experiments with $\spix=5\muK$ and compare
the performance of the lensing estimators in Fig.~\ref{fig:com}. 
Since the reconstruction becomes noisier in the
presence of detector noise and telescope beam,
10,000 clusters are stacked for the mean mass profiles when the standard
or modified quadratic estimator is used, while the improved quadratic estimator
is iteratively applied to only 1000 clusters. For clusters
of $M=5\times10^{14}\hmsun$ at $z_L=1$, Fig.~\ref{fig:com}$a$ shows that 
the standard 
quadratic estimators become substantially biased  in a region around
massive clusters, consistent with the previous results 
\citep{MABAMEET05,HUDEVA07}. Quadratic terms in $\phi_\vL$ ignored in
the linear approximation coherently contribute to $\khat_\vL$, and hence the
reconstructed $\khat(\Vang)$ is biased 
low where the linear approximation is violated \citep{HUDEVA07}.

Next we consider a modified quadratic estimator in Fig.~\ref{fig:com}$b$ 
and adopt $\lcut=1500$. The modified quadratic
estimator recovers the true mass profile within $\Rvir$ but with small
deviation beyond $\Rvir$. The modified quadratic estimators operate
in the same way of the standard quadratic estimators, except signals are
removed on small scales ($l\geq\lcut$), 
where the linear approximation is violated.
However, the choice of $\lcut$ is rather arbitrary and should be calibrated
against simulations: lower $\lcut$ is needed for more massive clusters.
Note that the modified quadratic estimator with 
$\lcut\rightarrow\infty$ 
exactly reduces to the standard quadratic estimator
(in practice $\lcut\gtrsim10^4$ can achieve this limit because of the Silk 
damping). In other words, a modified quadratic estimator 
with $\lcut\simeq10^4$ fails to reconstruct
the mass profile (born out by Fig.~\ref{fig:com}$a$).
Moreover, we had to adopt $\lcut=1500$ to reconstruct the mass profile
in Fig.~\ref{fig:com}$b$ and~\ref{fig:com}$d$, a more aggressive choice
than $\lcut=2000$ proposed in \citep{HUDEVA07}, with which
we cannot recover the mass profile.  This reflects the
sensitivity of the modified quadratic estimator to $\lcut$ as a function of
cluster mass.
Larger number of clusters are also required to reconstruct the true mean
mass profile due to the reduction in the signal-to-noise ratio.

Figure~\ref{fig:com}$c$ shows the reconstruction by our improved quadratic
estimator with $\Minit=1\times10^{14}\hmsun$. The improved quadratic 
estimator recovers
the true mass profile with no significant bias in the presence of detector
noise. After a few iterations, the estimates quickly converge to the true
model and the scatter around the mean is greatly reduced compared to 
Fig.~\ref{fig:com}$b$. Note that we iteratively applied the improved quadratic
estimator to the same 1000 clusters.

In Fig.~\ref{fig:com}$d$ and \ref{fig:com}$e$, we consider the effect of
telescope beam with $\tfwhm=0.\!'5$. Both estimators
in Fig.~\ref{fig:com}$d$ and \ref{fig:com}$e$ recover the true mass profile
unbiased in the presence of detector beam, while there exist some deviations
in both cases.
However, note that we explicitly account for the beam 
effect using the formulas developed in Sec.~\ref{ssec:qe}, rather than 
deconvolve the beam before applying the lensing estimators. The latter
approach often used in the literature suffers from deconvolved detector noise
exponentiating on small scales. This problem requires a low-pass filtering
of reconstructed $\khat(\Vang)$, additionally
removing the signals below the beam scale,
which results in a distortion of its shape of $\khat(\Vang)$,
making it hard to compare directly to theoretical predictions. 
However, in reality 
beam convolution suppresses detector noises (of course  lensing signals
as well), and it simply makes the reconstruction noisy below the beam scale.
The dot-dashed line in Fig.~\ref{fig:com}$d$
contrasts the reconstruction when we explicitly remove 
$\khat(\Vang)$ at $l\geq1/\sbeam$, where $\khat(\Vang)$ is
obtained by applying the modified quadratic estimator with
beam-deconvolved data (the line is displaced to avoid confusion with other
lines). Significant shape distortion in $\khat(\Vang)$ 
complicates the interpretation.

For a larger beam size comparable to the scale radius of the clusters
($\tfwhm\simeq1'$), the reconstruction becomes more challenging: modified
quadratic estimators cannot recover the cluster mass profile without 
significant shape distortion ({\it dot-dashed}).
The improved quadratic estimator in Fig.~\ref{fig:com}$f$
recovers the true mass profile beyond $\Rvir$, while it develops small
bias below the beam scale. 

\begin{figure}
\centerline{\psfig{file=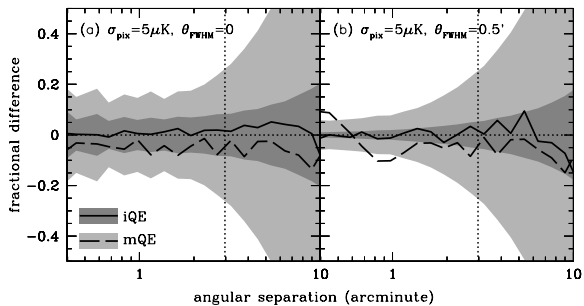, width=3.0in}}
\caption{Fractional difference between the lensing estimates and the true
cluster mass profile in Fig.~\ref{fig:com}. The difference ({\it lines})
is computed from the mean mass profiles obtained by stacking 10,000 clusters
for both estimators,
while the statistical uncertainty ({\it gray bands}) in the difference is 
scaled for 500 clusters. The vertical dotted lines show
the cluster virial radius.}
\label{fig:err}
\end{figure}

Figure~\ref{fig:err} plots the fractional difference between the lensing 
estimates and the true cluster mass profile in Fig.~\ref{fig:com}, 
comparing their uncertainty in the mean profile. The difference ({\it lines})
is computed from the mean mass profiles by stacking 10,000 clusters for 
both estimators, while the statistical uncertainty
({\it gray bands}) in the difference is scaled for 500 clusters for 
comparison.
The left panel shows that both estimators recover the cluster mass profile
at the 5\% level or better in the absence of telescope beam, 
while the modified quadratic estimator may need fine-tuning of $\lcut$ to
achieve better accuracy. However, the difference in their measurement 
uncertainty is in stark contrast: the improved quadratic estimator has a
significantly higher signal-to-noise ratio than the modified quadratic
estimator. While the reconstruction becomes harder especially beyond $\Rvir$
in the presence of telescope beam shown in the right panel,
this trend of signal-to-noise ratio difference persists.
Note that due to the beam smoothing effect the uncertainty in the estimates 
at $\theta\leq\tfwhm$ is reduced while it is highly correlated among adjacent
bins.

So far 
we have numerically demonstrated the performance of the lensing estimators
in Figs.~\ref{fig:com} and~\ref{fig:err}:
standard quadratic estimators are significantly 
biased; modified and improved quadratic estimators recover the cluster 
mass profile with no bias, while they show substantial difference
in the number of clusters that is required to obtain the mean mass profile.
To quantify this difference, we evaluate 
$\Delta\chi^2$ of each lensing estimator
\beeq
\Delta\chi^2=\sum_{\theta,\theta'}\kappa(\theta)~C_{\khat}^{-1}(\theta,\theta')
~\kappa(\theta'),
\eneq
where the covariance matrix of $\khat(\theta)$ is 
\beeq
C_{\khat}(\theta,\theta')=\left\langle\left[\khat(\theta)-\kappa(\theta)\right]
\left[\khat(\theta')-\kappa(\theta')\right]\right\rangle.
\eneq
Since $\khat(\Vang)$ is computed from the two Wiener-filtered functions of
the CMB temperature anisotropies, the covariance matrix is non-diagonal.
The finite width of the convolution filter $H(\Vang)$ in Eq.~(\ref{eq:filterH})
also reflects that the lensing estimators are a non-local function of the CMB
temperature anisotropies, and hence non-zero $C_{\khat}$ 
when $\theta\neq\theta'$.

In the absence of telescope beam in Figs.~\ref{fig:com}$b$,~\ref{fig:com}$c$,
and~\ref{fig:err}$a$, the ratio of $\Delta\chi^2$ for the modified quadratic
estimator relative to the improved quadratic estimator is 8.1:
a factor of eight larger number
of clusters is required for the modified quadratic estimator to achieve the
same level of accuracy than that for the improved quadratic estimator.
In the presence of telescope beam in Figs.~\ref{fig:com}$d$,~\ref{fig:com}$e$,
and~\ref{fig:err}$b$, 
beam smoothing substantially degrades the ability to recover the
true cluster mass profile for both estimators, and its effect is
relatively larger for the modified quadratic estimator, increasing
the ratio to 10.4. 

\subsection{Sunyaev-Zel'dovich Effects}
\label{ssec:sz}
On small scales ($l>2000$), the primordial CMB temperature anisotropies decay
exponentially due to the Silk damping \citep{SILK68} and the dominant source
of secondary anisotropies is the thermal Sunyaev-Zel'dovich (tSZ) effect,
arising from scattering off hot electrons in massive clusters. However, the
tSZ effect imprints a unique frequency dependence in the CMB temperature
anisotropies, which in principle can be used to remove the tSZ signals.
The same Compton scattering process also gives rise to a Doppler effect in the 
CMB temperature anisotropies due to the bulk motion of electron gas, or
the kinetic Sunyaev-Zel'dovich (kSZ) effect (see \citep{BIRKI99,CAHORE02}
for recent reviews). These kSZ signals, albeit smaller than tSZ signals,
are spectrally indistinguishable from the intrinsic CMB temperature 
anisotropies, introducing an artifact in the lensing reconstruction. Here
we assume that the tSZ signals can be cleaned perfectly, and 
we investigate how the kSZ signals deteriorate the lensing reconstruction.

For simplicity, we assume that the gas density traces the dark matter 
distribution in a massive cluster, with the same NFW profile. Given the 
line-of-sight velocity $v_\up{los}$ of the cluster, the kSZ effect results
in temperature anisotropies
\beeq
\Delta T(\theta)=-v_\up{los}~\tau(\theta)~T_\up{CMB}\equiv-\Delta T_\up{kSZ}
~{\Sigma(\theta)\over\Sigma(0)},
\label{eq:kszT}
\eneq
where $\tau(\theta)$ is the Thompson scattering
optical depth, proportional to the projected density $\Sigma(r=D_L~\theta)$. 
We parametrized the product of 
$v_\up{los}$ and $\tau(0)$ as $\Delta T_\up{kSZ}$. 
Note that since the intrinsic CMB and kSZ induced anisotropies dilute in the
same way as the universe expands, there is no $(1+z_L)$ factor in 
Eq.~(\ref{eq:kszT}) and $T_\up{CMB}=2.725$K is the CMB temperature today.

For a typical cluster with
electron number density $\sim0.01~\up{cm}^{-3}$ and core radius $\sim100$~kpc,
the Thompson scattering optical depth is $\tau(0)=2\times10^{-3}$ at the core. 
The rms velocity
dispersion in linear theory is $\sigma_v=1.3\times10^{-3}(=390~\kms)$ 
at $z_L=1$,
and this results in the rms temperature fluctuation $\Delta T_\up{kSZ}=3.7\muK$
at the core. We randomly draw $\Delta T(0)$ from a Gaussian distribution with
zero mean and dispersion $\sigma=\Delta T_\up{kSZ}$, 
then we add $\Delta T(\Vang)$ to $\lenT(\Vang)$ for observations of 
each cluster. 

\begin{figure}
\centerline{\psfig{file=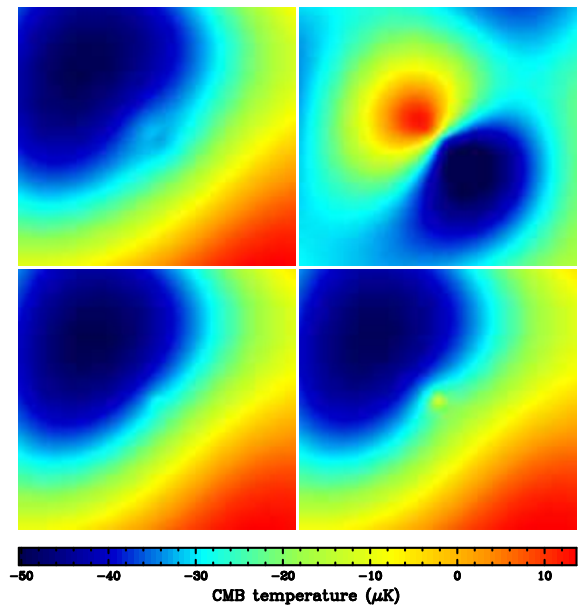, width=3.0in}}
\caption{(color online)
Cluster lensing and kinetic Sunyaev-Zel'dovich (kSZ) effects on 
the CMB. For comparison, we plot $6'\times6'$ regions of CMB temperature maps
around a cluster of $M=5\times10^{14}\hmsun$ ($\theta_\up{vir}=3.\!\!'0$) 
at $z_L=1$.
{\it Upper panels}: lensed temperature map ({\it left})
and its difference from the intrinsic temperature map ({\it right}).
{\it Bottom panels}: assuming that the cluster is 
moving toward an observer, the kSZ effect is set
$\Delta T_\up{kSZ}=3$ ({\it left}) and $15\muK$ ({\it right}) at the center.
The color scales in each panel represent the same temperature except in the 
upper right panel, where the color represents the difference ranging from
$-5\muK$ to $5\muK$.}
\label{fig:kszshow}
\end{figure}

First, we compare the cluster lensing and kSZ effects on the CMB temperature
field. Figure~\ref{fig:kszshow} plots a 6'$\times$6' regions of CMB maps
around a cluster of $M=5\times10^{14}\hmsun$ ($\theta_\up{vir}=3.\!\!'0$) 
at $z_L=1$.
The top panels show the lensed temperature field ({\it left}) and the
difference from the intrinsic temperature field ({\it right}). Gravitational
lensing imprints dipole-like wiggles in the CMB map on top of the smooth
large-scale gradient field. Perpendicular to the gradient direction there
exists no temperature change and hence lensing reconstruction is degenerate
along the direction. The bottom panels show the kSZ effect with
$\Delta T_\up{kSZ}=3\muK$ ({\it left}) and $15\muK$ ({\it right}). We assume
that the cluster is moving toward the observer. With the small optical depth
in the left panel, the kSZ effect is relatively small compared to the lensing
effect. Larger optical depth in the right panel substantially enhances the
kSZ effect, dominating over the lensing effect at the center. However, since
the lensing effect is much less concentrated than the kSZ effect as the 
dipole-like wiggles peak at a few scale radii ({\it top right}),
the reconstruction is still possible.

\begin{figure}
\centerline{\psfig{file=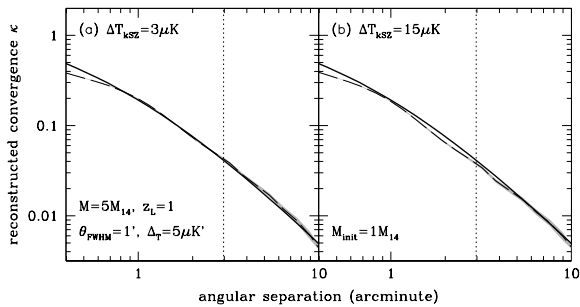, width=3.0in}}
\caption{Impact of kinetic Sunyaev-Zel'dovich (kSZ) effects on the mass profile
reconstruction. Assuming that the gas distribution traces
the dark matter distribution in clusters, the kSZ effect is computed by 
assigning a Gaussian random velocity to each cluster
with rms temperature change $\Delta T_\up{kSZ}=3$ ({\it left}) and 
$15\muK$ ({\it right}) at the center, respectively.}
\label{fig:ksz}
\end{figure}

Figure~\ref{fig:ksz} shows the impact of the kSZ effect on reconstructing
mass profiles. For clusters of $M=5\times10^{14}\hmsun$ at $z_L=1$ in 
an experiment
with $\tfwhm=1'$ and $\spix=5\muK$, we iteratively
use improved quadratic estimators with $\Minit=1\times10^{14}\hmsun$.
The mean and the uncertainties are computed
from 1000 clusters. Figure~\ref{fig:ksz}$a$ shows that the kSZ effect with
$\Delta T_\up{kSZ}=3\muK$ has relatively little impact on the reconstruction:
the kSZ effect becomes negligible beyond $r_s$ because the density profile
declines $r^{-3}$ (the gas density in reality would be steeper and
more confined to the center than we assumed here). The lensing effect, on the
other hand, is sensitive to the deflection field and remains strong beyond
$r_s$, declining less rapidly than the kSZ effect \citep{SEZA96}.
In Fig.~\ref{fig:ksz}$b$,
we consider a larger kSZ effect with $\Delta T_\up{kSZ}=15\muK$, expected
either from higher electron number density or from higher matter fluctuation
normalization $\rms\propto\sigma_v$. With the temperature anisotropies 
comparable to the lensing effect, the reconstruction becomes difficult and
it starts to develop bias around $\Rvir$ as $\Delta T_\up{kSZ}$ increases.
Note that the bias at the center is largely due to the telescope beam effect.

In the presence of contaminants such as residual foreground or tSZ effect,
radio point sources, and large kSZ effect, the lensing estimators based on
temperature anisotropies need to be complemented by using lensing estimators
based on combination of temperature and E- and B-mode polarization
\citep{OKHU03}, since there exists no significant source of contamination that
mimics the intrinsic CMB polarization. Furthermore, the unique relation between
the E- and B-mode polarization signals \citep{SEZA97,KAKOST97} can be used to
provide a robust consistency check. However, measurements of the lensed
polarization fields would require an experiment with higher angular resolution
and sensitive detectors than experiments that are currently available.

\section{Discussion}
\label{sec:con}
Weak gravitational lensing of the CMB gives rise to a deviation of the 
two-point correlation function of the CMB temperature anisotropies from 
otherwise statistically isotropic function. Quadratic estimators 
\citep{HU01b} have been
widely used to reconstruct cluster mass profiles and large-scale structure
by measuring the induced anisotropies in the two-point correlation function.
We have shown that standard quadratic estimators become optimal in the
limit of no lensing, saturating the Cram\'er-Rao bound, while they become
progressively biased and sub-optimal as the lensing effect increases. 
Especially for clusters that can be found by the ongoing SZ surveys like 
ACT and SPT, the standard quadratic estimators start to be biased at 
$z_L\simeq0.3$, and at higher redshift, where the lensing effect is larger,
other estimators should be used to reconstruct cluster mass profiles.

It is recently proposed \citep{HUDEVA07}
that this obstacle in the standard quadratic estimators
can be overcome by explicitly removing the signals in the CMB temperature
gradient field at $l\geq\lcut$, where the lensing effect is large in violation
of the linear approximation. However, although these
modified quadratic estimators
recover cluster mass profiles with no significant bias, the choice of $\lcut$
is somewhat arbitrary and it depends on the lensing effect, which requires
prior calibrations against numerical simulations before one can apply the
modified quadratic estimators to CMB maps.

We have developed a new maximum likelihood estimator for reconstructing 
cluster mass profiles and large-scale structure. We first construct a CMB
temperature field by delensing the observed temperature field based on an
assumed mass model. We have proved that the delensed temperature field is 
close to the unlensed temperature field with telescope beam smoothed and
detector noise added, if the assumed mass model is a good approximation to 
the true mass model. The delensed temperature field can then be used to set up
the likelihood of the CMB, and our new estimator that maximizes this 
likelihood takes a similar form of the standard quadratic estimators,
because it approaches to an optimal estimator as the assumed model becomes
the true model. Our maximum likelihood estimator can be iteratively applied
as we
update the assumed mass model, until it converges (to the true model) and
the estimate is consistent with the assumed model. Our maximum likelihood
estimator, named as an improved quadratic estimator, is easy to implement
in practice and it has no free parameter.

Our improved quadratic estimators quickly converge to the true mass model
after a few iterations, even when an assumed initial model is 
significantly different from the true model. 
When the estimate is inconsistent with the
assumed model, one can adopt another initial model for iterations for faster
convergence of the improved quadratic estimators. The telescope beam and 
detector noise renders the reconstruction harder, but we have demonstrated
that the improved quadratic estimators recover cluster mass profiles with a beam
size comparable to the cluster scale radius.
Furthermore, our new estimator significantly improves the signal-to-noise
ratio over the standard or modified quadratic estimators by a factor of 
ten in number of clusters, because when
an assumed model is close to the true mass model, the only source of noise
for our estimator is the intrinsic fluctuations of the CMB temperature 
gradient.

We have tested the robustness of the improved quadratic estimators in the
presence of the kSZ effect. The kSZ distortion $\Delta T_\up{kSZ}\leq15\muK$
at the center results in relatively small bias in the reconstructed cluster
mass profiles. However, since the optical depth is a function of 
electron number density in the clusters, it is related to the true mass
profile.
Therefore, we could take a more aggressive approach to
modeling kSZ signals from an assumed initial mass model and subtract the
kSZ contributions before applying  improved quadratic estimators. 
Furthermore, this template for kSZ signals can also be iteratively refined 
as we update our assumed mass model.

Since the reconstruction is non-parametric, it is not limited to spherical
clusters, while stacking many clusters ensures that irregular shapes of
individual clusters become irrelevant. 
Similar arguments can be applied to projection effects: 
each cluster can be located at a line-of-sight with overdense or underdense
regions, but projection effects become negligible once many lines-of-sight are 
combined.
Given a sample of clusters from SZ surveys, the average mass profile
of stacked clusters would provide a cluster-mass cross-correlation function,
which can be used to measure the growth rate of structure, probing the 
evolution of dark energy, instead of individual cluster mass profiles.

However, in reality it would be harder to reconstruct
cluster mass profiles than considered here, because there exist other
contaminants such as point radio sources and residual foreground and/or tSZ
effect, and other complications such as non-isolated clusters
and internal bulk motion of gas in clusters. However, additional information
from polarization measurements may be used to overcome some of the 
difficulties, given the unique relation between the E- and B-mode polarization
signals and relatively negligible primary and secondary contaminants.
Finally we mention that our improved quadratic estimators can be applied 
to reconstruct large-scale structure, while in this regime standard quadratic 
estimators can be used without significant bias.

\acknowledgments 
We thank Oliver Zahn for useful discussion. J.~Y. thanks C.~K.~Chan for 
technical help on FFTw routines. J.~Y. is supported by the Harvard College 
Observatory under the Donald~H. Menzel fund.
M.~Z. is supported by the David and Lucile Packard, the Alfred~P. Sloan,
and the John~D. and Catherine~T. MacArthur Foundations.
This work was further supported by NSF grant AST~05-06556 and NASA ATP
grant NNG~05GJ40G.

\appendix
\section{Delensed Temperature Field}
\label{app:delens}
Here we derive a relation 
$\That_\vL\simeq T_\vL ~e^{-{1\over2}l^2\sbeam^2}+T^N_\vL$ in the 
presence of telescope beam and detector noise. Given the lensing potential
$\phi^m(\Vang)$ of an assumed mass model, the lensing equation relates an image
position $\Vang$ to a source position $\bhat^m=\Vang+\ang\phi^m(\Vang)$.
Here we keep the superscript $m$ to indicate the relation to the assumed
model. The true source position is then
$\bhat=\Vang+\ang\phi(\Vang)$, where $\phi(\Vang)$ is the true lensing 
potential. Now we construct a delensed temperature field
\bear
\That(\bhat^m)&=&\Tobs(\Vang) \\
&=&\int d^2\Vangm ~B(\Vangm-\Vang)~\lenT(\Vangm)+T^N(\Vang), \nonumber
\enar
where $B(\Vangm)$ is the telescope beam function.
Since the lensing equation is not analytically invertible in general, we
keep both $\bhat^m$ and $\Vang$, but note that they are not independent
variables. In Fourier space, the delensed temperature field is
\beeq
\That_\vL=\int d^2\bhat^m ~\That(\bhat^m)~e^{-i\vL\cdot\bhat^m}\equiv
\That^S_\vL+\That^N_\vL,
\eneq
with a contribution from the CMB
\bear
\That^S_\vL&=&\int d^2\bhat^m \int d^2\Vangm~ B(\Vangm-\Vang)~\lenT(\Vangm)
~e^{-i\vL\cdot\bhat^m} \nonumber \\
&=&\int d^2\vL_1~ B_{\vL_1}\lenT_{\vL_1}\int{d^2\bhat^m\over(2\pi)^2}~
e^{i\vL_1\cdot\Vang-i\vL\cdot\bhat^m},
\label{app:gen}
\enar
and a contribution from the detector noise
\bear
\That^N_\vL&=&\int d^2\bhat^m ~T^N(\Vang)~e^{-i\vL\cdot\bhat^m}\\
&=&\int d^2\vL_1~ T^N_{\vL_1}\int{d^2\bhat^m\over(2\pi)^2}~
e^{i\vL_1\cdot\Vang-i\vL\cdot\bhat^m}. \nonumber
\enar

\begin{figure}
\centerline{\psfig{file=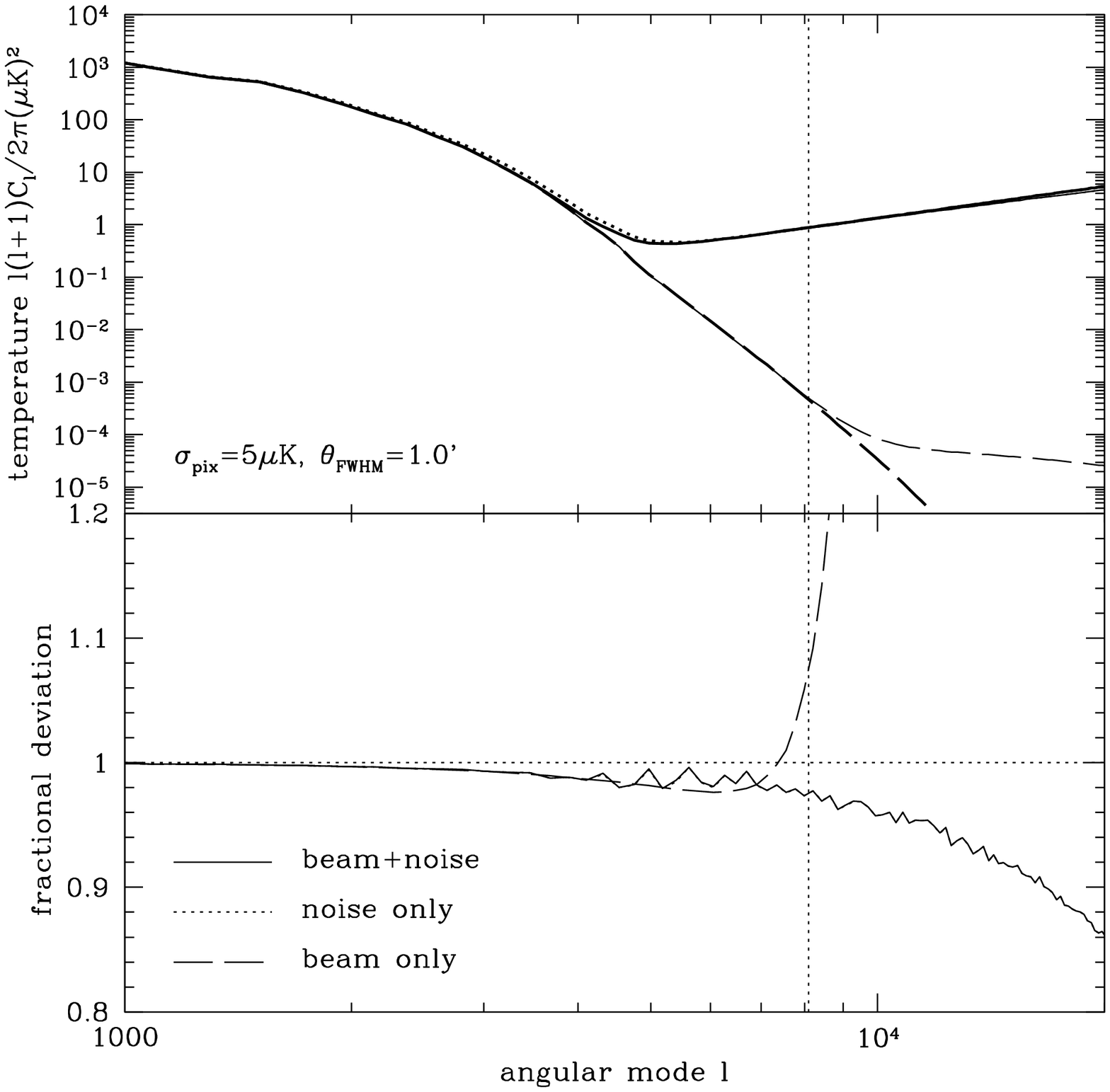, width=3.0in}}
\caption{Effects of telescope beam and detector noise on the delensing 
process. The top panel compares $\That_\vL$ ({\it thin}) with 
$T_\vL ~e^{-{1\over2}l^2\sbeam^2}+T^N_\vL$ ({\it thick}) in terms of their 
power spectrum, and the bottom panel shows the fractional deviations.
The vertical dotted line represents the beam scale $l=1/\sbeam$.
CMB experiments
with $\tfwhm=1'$ and $\spix=5\muK$ are considered for clusters of 
$M=5\times10^{14}\hmsun$ at $z_L=1$. The noise only case is largely obscured by 
the solid line.}
\label{fig:app}
\end{figure}

The lensed temperature is $\lenT(\Vang)=T(\bhat)$ and its Fourier mode is
\beeq
\lenT_\vL=\int d^2\vL_1~T_{\vL_1}\int{d^2\Vang\over(2\pi)^2}~
e^{-i\vL\cdot\Vang+i\vL_1\cdot\bhat}.
\eneq
With the linear approximation, one can expand the exponential term to the
first order in $\phi_\vL$ and this equation reduces to Eq.~(\ref{eq:lensexp}).
However, we keep the equation as general as possible to be valid, even when
the lensing effect is large. Substituting $\lenT_{\vL_1}$ in Eq.~(\ref{app:gen})
and changing the integration variable $\Vang$ to $\bhat^m$ gives
\bear
\That^S_\vL&=&\int d^2\vL_1\int d^2\vL_2~B_{\vL_1}T_{\vL_2}
\int{d^2\Vang\over(2\pi)^2}\int{d^2\Vang_2\over(2\pi)^2}
\left|{d^2\bhat^m\over d\Vang^2}\right|\nonumber \\
&\times&e^{i\vL_1\cdot(\Vang-\Vang_2)}e^{i(\vL_2\cdot\Vang_2-\vL\cdot\Vang)}~
e^{i\left[\vL_2\cdot\ang\phi(\Vang_2)-\vL\cdot\ang\phi^m(\Vang)\right]}.
\enar
Given the lensing potential $\phi(\Vang)$ (analogously for $\phi^m(\Vang)$),
the Jacobian is related to the distortion matrix 
\beeq
\left|{d^2\bhat\over d\Vang^2}\right|=\left|\bdv{M}^{-1}\right|=
\left|\bdv{I}+\ang\ang\phi\right|=
\left|\left[1-\kappa(\Vang)\right]^2-\gamma^2(\Vang)\right|,
\eneq
and its inverse is the lensing magnification.

For a Gaussian beam $B_\vL=\exp\left[-{1\over2}l^2\sbeam^2\right]$, we can
integrate over the beam factor
\bear
\That^S_\vL&=&\int d^2\vL_2~T_{\vL_2}\int{d^2\Vang\over(2\pi)^2}
\int{d^2\Vang_2\over(2\pi)^2}\left|{d^2\bhat^m\over d\Vang^2}\right| \\
&\times&
{2\pi\over\sbeam^2}e^{-{|\Vang-\Vang_2|^2\over2\sbeam^2}}
e^{i(\vL_2\cdot\Vang_2-\vL\cdot\Vang)}~
e^{i\left[\vL_2\cdot\ang\phi(\Vang_2)-\vL\cdot\ang\phi^m(\Vang)\right]}. \nonumber
\enar
Now we parametrize $\Vang_2$ by a dimensionless displacement vector
$\dsmall$ centered at $\Vang$ (i.e., $\Vang_2=\Vang+\sbeam\dsmall$).
The Gaussian beam factor guarantees that the integrand is non-vanishing only
when $\Delta=|\dsmall|$ is small. In order to get more intuition, we expand
$\phi(\Vang_2)\simeq\phi(\Vang)+\ang\phi(\Vang)\cdot\sbeam\dsmall$
to the linear order in $\Delta$, and integrating over $\dsmall$ gives
\bear
\That^S_\vL&=&\int d^2\vL_2~T_{\vL_2}\int{d^2\Vang\over(2\pi)^2}
\left|{d^2\bhat^m\over d\Vang^2}\right|\\
&\times&e^{i(\vL_2\cdot\bhat-\vL\cdot\bhat^m)}~
e^{-{1\over2}\sbeam^2|\bdv{M}^{-1}\cdot\vL_2|^2}. \nonumber
\enar
This is the final expression for the delensed temperature field. The first
exponential
term of the integrand controls the delensing process: when the assumed model
is close to the true model after a few iterations 
($\phi^m(\Vang)\simeq\phi(\Vang)$, $\bhat^m\simeq\bhat$), the integral becomes
a Dirac delta function and $\That^S_\vL=T_\vL$, when the beam smoothing is
negligible. The distortion matrix is close to the identity matrix beyond
$\Rvir$ and $\That^S_\vL\simeq T_\vL ~e^{-{1\over2}l^2\sbeam^2}$. Around massive
clusters, the distortion matrix deviates from the identity matrix and its
determinant becomes smaller than one, making the exponential factor unity.
This reflects that the beam size is reduced by a mapping from the image
plane to the source plane, and practically 
$\That^S_\vL\simeq T_\vL~ e^{-{1\over2}l^2\tilde\sbeam^2}$ with
$\tilde\sbeam<\sbeam$.

For a white detector noise, the delensed detector noise is simply the
redistributed white noise. However, since the delensing process alters the unit
area on the sky,
it becomes non-white but its deviation is confined to relatively small
region; the noise power spectrum is
\bear
\langle\That^N_\vL\That^{N*}_{\vL'}\rangle&=&
\int d^2\vL_1\int d^2\vL_2~\langle T^N_{\vL_1}T^{N*}_{\vL_2}\rangle \nonumber \\
&\times&\int{d^2\bhat^m_1\over(2\pi)^2}\int{d^2\bhat^m_2\over(2\pi)^2}~
e^{i\vL_1\cdot\Vang_1-i\vL\cdot\bhat^m_1}~e^{-i\vL_2\cdot\Vang_2+i\vL'\cdot\bhat^m_2}
\nonumber\\
&=&C^N\int{d^2\bhat^m_1}\int{d^2\bhat^m_2}~\delta(\Vang_1-\Vang_2)~
e^{-i\vL\cdot\bhat^m_1+i\vL'\cdot\bhat^m_2} \nonumber \\
&=&C^N\int{d^2\bhat^m_1}\left|{d^2\bhat^m_1\over d\Vang_1^2}\right|
e^{-i(\vL-\vL')\cdot\bhat^m_1}.
\enar
It is the Jacobian of the distortion matrix that 
makes white noise non-white in a region around massive clusters. Outside
$\Rvir$, where the Jacobian is near unity, the integral becomes a Dirac
delta function and the noise is again white.

Figure~\ref{fig:app} compares our delensing ($\That_\vL$: {\it thin})
and perfect delensing ($T_\vL ~e^{-{1\over2}l^2\sbeam^2}+T^N_\vL$: {\it thick})
processes in terms of their power spectrum. In the absence of detector noise
({\it dashed}), $\That^S_\vL$ starts to deviate from 
$T_\vL~ e^{-{1\over2}l^2\sbeam^2}$ around the beam scale $l\simeq1/\sbeam$
({\it vertical dotted}), declining less rapidly. On scales below the beam scale,
our approximation ($\Delta\ll1$) breaks down and $\bdv{M}^{-1}(\Vang)$ differs
from the identity matrix, leading to the excess power. 
However, at this scale, signals are dominated by the detector noise
({\it solid}). Since detector noises are unaffected by the beam distortion
when delensed, the deviation of $\That^N_\vL$ from $T^N_\vL$
is relatively mild and it is solely due to the (inverse) magnification effect
of the mapping from the image plane to the source plane. The noise only
case ({\it dotted}) is largely obscured by the solid line. In summary,
telescope beam and detector noise has little impact on our delensing
process at scales larger than the beam scale, where most of the information
is contained.

\end{document}